\documentclass[aps,prl,letterpaper,showpacs,twocolumn,preprintnumbers,amsmath,amssymb,nofootinbib]{revtex4}
\usepackage{graphicx}% Include figure files
\DeclareMathAlphabet   {\mathsc}{OT1}{cmr}{m}{sc}

\def\[{\left [}
\def\]{\right ]}
\def\({\left (}
\def\){\right )}

\newcommand{\GeV}      {~\mathrm{GeV}}
\newcommand{\TeV}      {~\mathrm{TeV}}

\newcommand{\UV}       {\mathsc{uv}}

\newcommand{\GUT}      {\mathsc{gut}}

\newcommand{\gappeq}{\mathrel{\rlap {\raise.5ex\hbox{$>$}}
{\lower.5ex\hbox{$\sim$}}}}
\newcommand{\lappeq}{\mathrel{\rlap{\raise.5ex\hbox{$<$}}
{\lower.5ex\hbox{$\sim$}}}}

\begin{document}

\preprint{MCTP-03-16}

\title{Phenomenology and Theory of Possible Light Higgs Bosons}

\author{G.~L.~Kane}
\author{Brent D. Nelson}
\author{Ting T. Wang}
\affiliation{Michigan Center for Theoretical Physics,
University of Michigan, Ann Arbor, MI 48109, USA}

\author{Lian-Tao Wang}
\affiliation{Department of Physics, University of Wisconsin,
Madison, WI 53706, USA}

\date{\today}

\begin{abstract}
We study the implications of the absence of a direct discovery of
a Higgs boson at LEP. First we exhibit 15 physically different
ways in which one or more Higgs bosons lighter than the LEP limit
could still exist. In the minimal supersymmetric standard model
(MSSM) all of these, as well as the cases where $m_{h}\geq 115
\GeV$, seem fine-tuned. We examine some interpretations of the
fine tuning in high scale theories. The least fine-tuned MSSM
outcome will have $m_h \simeq 115\GeV$, while approaches that
extend the MSSM at the weak scale can naturally have larger $m_h$.
\end{abstract}

\pacs{14.80.Cp,12.60.Jv}
%\keywords{}

\maketitle

The Minimal Supersymmetric Standard Model is defined as the
simplest supersymmetric extension of the Standard Model (SM).
Every SM particle has a superpartner, the basic Lagrangian is
supersymmetric, and the gauge group is the same $SU(3)\times
SU(2)\times U(1)$ as that of the SM. The full supersymmetry is
softly broken by certain dimension two and three operators. There
is considerable indirect evidence that this theory is likely to be
part of the description of nature. If it is, a Higgs boson with
mass less than about $130\GeV$ must exist, and superpartners must
be found with masses not too much larger than those of the W, Z
and top quark. This theory has three neutral Higgs boson physical
states (denoted h, H, and A in the simplest CP conserving case)
and a charged Higgs pair H$^{\pm }$. Of course it could be an
extension of the MSSM that describes nature at the weak scale, but
most extensions behave similarly to the MSSM at low energies.

While the Higgs boson mass can be as heavy as $130\GeV$ in the
MSSM, it has been known for some time that most models imply a
lighter state, usually below about $110\GeV$, when constraints
from non-observation of superpartners (real or virtual) are
imposed, and including the constraint that the indirect arguments
for supersymmetry are valid without fine-tuning. Even allowing
that perhaps LEP has seen a Higgs boson with $m_{h}=115\GeV$,
there is some fine-tuning present because the tree level Higgs
mass is bounded by $M_{Z}$ so the one loop corrections have to
supply about $\delta m^2_h \simeq (70 \GeV)^2$ when added in
quadrature. That is possible but requires some parameters to be
larger than naively expected. Of course, discussing an issue such
as fine-tuning can only be done in the context of a theory. Fine
tuning in the low scale theory could be a clue to a natural
structure of the high scale theory. Or it could be a clue that the
low scale theory is effectively excluded. If the MSSM is extended
to have a larger gauge group in such a way that new scalars mix
significantly with the MSSM scalars then the tree level mass of
the lightest Higgs can be larger than $M_{Z}$ and no fine tuning
is needed to have a heavier Higgs spectrum.

It is interesting to ask if the Higgs sector of the MSSM could be
such that LEP would not have found a signal because of reduced
Higgs production cross sections there, or reduced branching
ratios. In the following we present in Table~\ref{tbl:signals} 15
physically different scenarios where that is indeed the case, plus
the minimal scenario with $m_{h}=115\GeV$~(No.~10), and the case
with all Higgs heavier than about $115 \GeV$~(No.~17). All are
allowed by other data. All are detectable at the Fermilab Tevatron
collider with sufficient luminosity. All satisfy the constraints
for electroweak symmetry breaking, though sometimes in
unconventional ways. Having found these we can then ask if any of
them seem to be less fine-tuned, and thus may point towards an
underlying theory.

%---------------------THE TABLE-----------------------------------------
\begin{table*}[tb] \caption{\label{tbl:signals} \textbf{Possible
explanations consistent with LEP Higgs search results.} Ranges of
neutral and charged Higgs masses consistent with background only
hypotheses as well as one, two or three ``signal'' hypotheses are
listed. The column headed by ``Signals'' indicates what signals
might have appeared for a given model. Qualitative $\tan\beta$ and
Higgs coupling ranges for each individual parameter space is
given. All ranges should be understood as indicative of the
allowed region at the roughly 10\% accuracy level: fine scans of
the parameter space have not been performed. For Higgs state
$\varphi _{i}$ the $ZZ\varphi _{i}$ coupling is
$(g_{2}M_{Z}/\cos\theta _{W}) C_{i}$, approximate values are given
in the table. The column marked $\phi$ indicates a non-trivial
phase $\phi_{\mu A_t}$ is needed. When there is nontrivial phase,
$m_A$ is understood as the mass of the neutral higgs with smallest
$C_{ZZH_i}$ coupling. The column $\mu$ indicates the presence of a
large $\mu$ term. The column marked U indicates this scenario is
compatible with a unified SUSY breaking scenario such as mSUGRA.
We think all other such scenarios effectively reduce to one of
these.}
\begin{tabular}{|c||c|c|c|c|c|c|c|c|c|c|c|} \hline
No. & \parbox{1.9cm}{$m_h$} & \parbox{1.9cm}{$m_A$} &
\parbox{1.9cm}{$m_H$} & \parbox{1.9cm}{$m_{H^\pm}$} &
\parbox{2cm}{Signals}
& \parbox{1.5cm}{$\tan\beta$} & \parbox{1.5cm}{$C^2_{h}$} &
\parbox{1.5cm}{$C^2_{H}$} & \parbox{0.5cm}{U} & \parbox{0.5cm}{$\mu$} &
\parbox{0.5cm}{$\phi$} \\ \hline \hline
1 & 98 & 89 & 115 & 112-123 & 98,115,187 & 6-12 & 0.2 & 0.8 & & Y
& Y \\ \hline
2 & 98 & $<m_h$ & 115 & 106-127 & 98,115 & 4-13 & 0.2 & 0.8 & & Y
& Y \\ \hline
3 & 98 & $\approx m_h$ & 115 & 121-136 & 98,115 & 5-50 & 0.2 & 0.8
& & & \\ \hline
4 & 98 & 115-130 & 115 & 112-124 & 98,115 & 10-24 & 0.2 & 0.8 & &
Y & \\ \hline
5 & 70-91 & 96-116 & 115 & 110-140 & 115,187 & 10-50 & 0.0 & 1.0 &
& Y & \\ \hline
6 & 98 & 89 & $>115$ & 118-127 & 98,187 & 6-10 & 0.2 & 0.8 & & Y &
Y
\\ \hline
7 & 82-110 & $<m_h$ & 115 & $\sim m_A$ & 115 & 7-50 & 0.0 & 1.0 &
& Y & Y \\ \hline
8 & 82-110 & $\approx m_h$ & 115 & $\sim m_A$ & $115^{c}$ & 5-50 &
0.0 & 1.0 & Y & & \\ \hline
9 & 82-110 & 115-140 & 115 & $\sim m_A$ & 115 & 6-24 & 0.0 & 1.0 &
& Y & \\ \hline
10 & 115 & \multicolumn{2}{|c|}{$m_A\approx m_H>115$} & $\sim m_A$
& $115^{c}$ & 3-50 & 1.0 & 0.0 & Y & & \\ \hline
11 & 98 & 100-130 & 120-130 & $\sim m_A$ & 98 & 5-50 & 0.20 & 0.80
& & & \\ \hline
12 & 98 & $<98$ & 120-130 & 106-128 & 98 & 4-13 & 0.20 & 0.80 & &
Y & Y \\ \hline
13 & 65-93 & 94-120 & 116-125 & 110-140 & 187 & 8-50 & 0.0 & 1.0 &
& Y & \\ \hline
14 & 80-100 & 25-40 & 133-154 & 109-130 & None$^{a}$ & 2-5 &
0.5-0.8 & 0.2-0.5 & & Y & Y \\ \hline
15 & 111-114.4 &\multicolumn{2}{|c|}{$m_A\approx m_H>114.4$} &
$\sim m_A$ & None$^{b}$ & 2.4-4 & 1.0 & 0.0 & & & \\ \hline
16 & 70-114.4 & 90-140 & $>114.4$ & $\sim m_A$ & None & 4-50 & 0.0
& 1.0 & Y & & \\ \hline
17 & $>114.4$ & \multicolumn{2}{|c|}{$m_A\approx m_H>114.4$} &
$\sim m_A$ & None$^{c}$ & 4-50 & 1.0 & 0.0 & Y & & \\ \hline
\end{tabular}
\begin{flushleft} $^{a}$ Dominant decay is CP violating process $H_2 \to
H_1 H_1$. This case was studied in Ref.~\onlinecite{CaElPiWa00}.\\
$^{b}$ The ``invisible'' decay $h \to \tilde N_1\tilde N_1$ and $h
\to b \bar{b}$ decays are comparable. \\ $^{c}$ These scenarios
were studied in Ref.~\onlinecite{AmDeHeSuWe02}.
\end{flushleft}
\end{table*}
%------------------------------END OF THE TABLE------------------------

The ranges of masses in Table~\ref{tbl:signals} are representative
ones since we think more detailed study is not appropriate at this
stage. In some cases there could be two or even three light Higgs.
LEP groups have reported as many as three possible signals (each
about $2\sigma$)~\cite{LHWG01,LHWG02}, so for simplicity we
identify possible signals as the masses where the reported LEP
hints occur. We emphasize that we are not in any way claiming LEP
signals exist -- we are simply using these points to illustrate
the opportunities. If there were observable light Higgs bosons,
they are more likely to have masses where tentative signals have
been recorded.  We group the models according to whether three,
two, one or none of the light states could have given a signal at
LEP. We assume that if they did give a signal it was at one of the
$2\sigma$ reported masses $98 \GeV$ or $115 \GeV$ for one of the
two CP-even eigenstates, or where the sum of two masses $m_h + m_A
=187 \GeV$~\cite{So02a,So02b}. Three of the 16 light Higgs cases
have parameters that can exist in the unified minimal supergravity
(mSUGRA) paradigm; the others do not. Some require complex soft
terms. We also give low-scale parameters for some cases.

Because the one loop top/stop radiative correction to the Higgs
potential is rather large, a large phase (specifically the
relative phase of $\mu$ and $A_{t}$) can enter, and lead to a
relative phase between the Higgs vevs at the minimum of the
potential. This phase is physical and cannot be rotated away. It
leads to mixing between the mass eigenstates, and affects the
production rates and decay branching
ratios~\cite{BrKa98,KaWa00,CaElPiWa00,CaElPiWa02,CaElMrPiWa02}.
The column headed by $\phi$ has a Y if a non-trivial phase (not
zero or $\pi$) plays a role for a given model. When CP is
conserved we call the states by the usual names h, H and~A; one
can show here that $m_{h}$ is always less than ${\rm
max}(M_Z,m_{A})$, even allowing one-loop corrections for $m_h$, so
any model with $m_{A}$ and $M_{Z}<m_{h}$ requires a non-trivial
phase. This conclusion does not include loop effects for $m_A$, so
one can have $m_A$ a few GeV less than $m_h$ for certain
parameters if $\tan\beta$ is large. The column headed by $\mu$ has
a Y if $\mu$ is very large, say well above several hundred GeV;
this is particularly relevant because of the question of
fine-tuning needed to obtain electroweak symmetry breaking.

In most cases the charged Higgs mass $m_{H^{\pm}}$ is less than
the top quark mass, so the decay $t\rightarrow b+H^{\pm }$ is
allowed. Existing data from D0 excludes $m_{H^{\pm }}$ below about
$125\GeV$ for $\tan\beta$ larger than about 50 with mild model
dependence, so no model is fully excluded -- though parts of the
parameter range of some models are probably excluded by
non-observation of $H^{\pm }$. With more and better data from Run
II the $H^{\pm}$ of most of these models could be observed or
excluded~\cite{Dzero}. These small values for $m_{H^{\pm}}$ can
also exceed limits from ${\rm Br}(b \to s \gamma)$, but using
light chargino and gluino contributions provides significant
flexibility. However, cases~8 and~16 exceed the limits on ${\rm
Br}(b \to s \gamma)$ by more than a factor of two and are thus
likely to be excluded, though we should note that this is based on
using a unified mSUGRA model for these cases and may not hold when
departures from universality are entertained.

For completeness we exhibit a set of low-energy parameters that
determine the resulting Higgs sector of the model for three
entries in Table~\ref{tbl:signals}; in a longer paper we will
provide this information for all of them and give more details on
the ranges~\cite{future}.
%================= The three sample EW-scale cases ============
Entry No.~1 has $m_{h}=98\GeV$, $m_{H}=115\GeV$ and
$m_{h}+m_{A}=187\GeV$. Its parameters are $\tan\beta=6$, $|\mu|
=1700$, $|A_{t}|=|A_{b}|=400$, $\phi_{\mu}=-130^{o}$, $M_{1}=100$,
$M_{2}=200$, $M_{3}=600$ and
$m_{\tilde{Q}_3}=m_{\tilde{b}_{R}}=m_{\tilde{t}_{R}}=500$, with
all parameters in GeV. This gives $m_{H^{\pm }}=120\GeV$. The
masses of the three mass eigenstates are $m_1 =89.1$, $m_2 =97.4$,
and $m_3=115.0\GeV$, with $C_{i}^{2}$ respectively of 0.024,
0.229, and 0.747. All three states have ${\rm BR}(\varphi_{i} \to
b\bar{b}) \approx 0.91$. These give about $2\sigma$ signals at 98
and 115 GeV. Since $m_{A}\approx M_{Z}$ the $Zh$ and $Ah$ channels
add to give an apparent $187\GeV$ signal.

Entry No.~8 has $m_{H}=115\GeV$ with the other neutral Higgs
states having smaller masses. Its parameters are $\tan\beta=46.9$,
$\mu = -540$, $A_{t}=-758$, $A_{b}=-882$, $M_{1}=188$,
$M_{2}=351$, $M_{3}=1015$, $m_{\tilde{Q}_3}= 860$,
$m_{\tilde{b}_{R}}= 848$ and $m_{\tilde{t}_{R}}=776$, with all
masses in GeV. This gives $m_{H^{\pm }}=113.5\GeV$. The masses of
the three mass eigenstates are $m_h = 82.7$, $m_A =86.2$ and $m_H
= 114.7 \GeV$, with $C_{i}^{2}$ respectively of 0.013, 0 and
0.987. All three states have ${\rm BR}(\varphi_{i} \to b\bar{b})
\approx 0.934$ which yields an apparent $2\sigma$ signal at
$115\GeV$.

Entry No.~15 has no signal at LEP and a lightest Higgs boson mass
below $115\GeV$. Its parameters are $\tan\beta=2.4$, $\mu =190$,
$A_{t}=A_{b}=4000$, $M_{1}=55$, $M_{2}=250$, $M_{3}=700$ and
$m_{\tilde{Q}_3}=m_{\tilde{b}_{R}}=m_{\tilde{t}_{R}}=2000$, with
all masses in GeV. This gives $m_{H^{\pm }}=505\GeV$. The masses
of the three mass eigenstates are $m_h = 111.2$, $m_A =499.6$ and
$m_H = 504.3 \GeV$, with $C_{i}^{2}$ respectively of 0.999, 0 and
0.001. The branching ratios of the lightest state are ${\rm BR}(h
\to b\bar{b}) =0.296$ and  ${\rm BR}(h \to \tilde{N}_1
\tilde{N}_1) =0.621$, where $\tilde{N}_1$ is the stable lightest
superpartner and is a good candidate for the cold dark matter of
the universe. In the case presented here, $m_{\tilde{N}_1} = 43.5
\GeV$.
%======================== End of the three cases ==============

It would be very nice if one or more of the models pointed to a
simple high scale model which we could then study and perhaps
motivate. Unfortunately, this does not seem to occur. The low
energy values given above do not completely specify the MSSM soft
Lagrangian. Thus, translating these values to a high energy
boundary condition scale $\Lambda_{\UV}$ through renormalization
group (RG) evolution will involve some arbitrariness -- for
example, in choosing the low-energy values of slepton and
second-generation squark masses. In some instances, such as entry
No.~8 described above, the necessary low scale values could be
obtained from a unified mSUGRA model at the high scale. For the
remaining cases we have chosen representative points in the
low-energy allowed parameter space and evolved them from the
bottom upwards. Some subset of these results for the soft terms at
$\Lambda_{\UV} = \Lambda_{\GUT} = 1.9 \times 10^{16}$ are given in
Table~\ref{tbl:softhigh}, including those of Entries~1, 8 and~15.

Naturally, entries such as Nos.~8 and~16 which can be identified
with a point in the mSUGRA parameter space have a simple
appearance at the high scale. By contrast, those models which have
no mapping to a unified-mass model show no discernable pattern in
the soft Lagrangian. While some small degree of improvement may be
possible by varying those parameters left unspecified at the low
scale by the Higgs sector, we have found no instances where the
patterns of severe hierarchies and negative scalar mass-squareds
can be alleviated. Note that these non-universal cases are
particularly perverse in that both charge and color symmetries are
radiatively {\em restored} in these models as the parameters are
evolved towards the electroweak scale.

%===================== Soft term table ==============================
\begin{table}[tb] \caption{\label{tbl:softhigh} Relevant soft term and $\mu$ values at
the GUT scale for selected entries in Table~\ref{tbl:signals}.}
\begin{ruledtabular}
\begin{tabular}{|c|c|c|c|c|c|c|} \hline
Entry & 1 & 3 & 4 \& 9 & 8 & 15 & 16 \\
\hline \hline
$\tan\beta$ & 6 & 10 & 11.3 & 46.9 & 2.4 & 46.41 \\
\cline{1-7}
$M_{1}$ & 242 & 291 & 726 & 450 & 133 & 560 \\
$M_{2}$ & 243 & 292 & 365 & 450 & 304 & 560 \\
$M_{3}$ & 210 & 245 & 349 & 450 & 245 & 560 \\
\cline{1-7}
$A_t$ & 3266 & 3596 & -3835 & 0 & 26288 & 0 \\
$A_b$ & 1577 & 1799 & -840 & 0 & 8535 & 0 \\
\cline{1-7}
$m_{Q_3}^{2}$ & $-(529)^2$ & $(935)^2$ & -$(1018)^2$ & $(450)^2$ &
$(10059)^2$ & $(300)^2$ \\
$m_{U_3}^{2}$ & $-(682)^2$ & $(1397)^2$ & -$(1225)^2$ & $(450)^2$
& $(14101)^2$ & $(300)^2$ \\
$m_{D_3}^{2}$ & $-(196)^2$ & $-(355)^2$ & -$(759)^2$ & $(450)^2$ &
$(1908)^2$ & $(300)^2$ \\
%
%$m_{L_3}^{2}$ & $(470)^2$ & $(454)^2$ & $(510)^2$ & $(450)^2$ &
%$(1989)^2$ & $(300)^2$ \\
%
%$m_{E_3}^{2}$ & $(487)^2$ & $(486)^2$ & $(558)^2$ & $(450)^2$ &
%$(1999)^2$ & $(300)^2$ \\
%
\cline{1-7}
\, $m_{Q_{1,2}}^{2}$ & $-(246)^2$ & $-(417)^2$ & -$(772)^2$ &
$(450)^2$ & $(1890)^2$ & $(300)^2$ \\
\, $m_{U_{1,2}}^{2}$ & $-(180)^2$ & $-(371)^2$ & -$(756)^2$ &
$(450)^2$ & $(1902)^2$ & $(300)^2$ \\
\, $m_{D_{1,2}}^{2}$ & $-(182)^2$ & $-(366)^2$ & -$(732)^2$ &
$(450)^2$ & $(1902)^2$ & $(300)^2$ \\
%
%$m_{L_{1,2}}^{2}$ & $(470)^2$ & $(454)^2$ & $(510)^2$ &
%$(450)^2$
%& $(1989)^2$ & $(300)^2$ \\
%
%$m_{E_{1,2}}^{2}$ & $(487)^2$ & $(486)^2$ & $(558)^2$ & $(450)^2$
%& $(1998)^2$ & $(300)^2$ \\
%
\cline{1-7}
$m_{H_u}^{2}$ & $-(1880)^2$ & $(1684)^2$ & -$(2295)^2$ & $(450)^2$
& $(17114)^2$ & $(300)^2$ \\
$m_{H_d}^{2}$ & $-(1690)^2$ & $-(525)^2$ & -$(1989)^2$ & $(450)^2$
& $(412)^2$ & $(300)^2$ \\
\cline{1-7}
$\mu$ & -1687 & -492 & 1971 & -660 & 212 & -794 \\
\cline{1-7}
\end{tabular}
\end{ruledtabular}
\end{table}
%===============================================================

Even allowing for the possibility that some of the high-scale
values in Table~\ref{tbl:softhigh} which appear similar can, in
fact, be made to unify with the appropriate adjustment of low
scale values, we are still confronted with a large number of
unrelated parameters in the soft Lagrangian. Most models of
supersymmetry breaking (such as mSUGRA) are studied for their
simplicity; they tend to involve very few free parameters. The
traditional models of minimal gravity, minimal gauge and minimal
anomaly mediation, as studied in the Snowmass Points and
Slopes~\cite{Snowmass,DeHeSuWe03} have too few parameters to
possibly describe these nonuniversal cases {\em even when all
three are combined in arbitrary amounts}. Nor do string-based
models generally provide sufficient flexibility, whether they be
heterotic based~\cite{bench} or intersecting brane constructions
such as Type~IIB orientifold models~\cite{IbMuRi99}. While having
sufficient free parameters in the model is, strictly speaking,
neither necessary nor sufficient to potentially generate one of
the entries in Table~\ref{tbl:signals}, we feel it is a good
indication of the theoretical challenge faced by models that
cannot come from mSUGRA or other simple benchmark models. This is
particularly true when the number of free parameters within, say,
the scalar sector and the number of hierarchies in the soft
Lagrangian are considered.

That many of the entries in Table~\ref{tbl:signals} imply high
scale soft supersymmetry breaking patterns with such unattractive
features (and no discernable theoretical structure) can be
considered one element of the fine-tuning in such cases. In
Table~\ref{tbl:tuning} we have counted the number of apparent free
parameters for these sample cases, allowing for a generous
interpretation of which parameters could be made to unify. It is
not an automatic corollary, however, that the models that admit a
unified explanation are necessarily less fine-tuned.  In
Table~\ref{tbl:tuning} we also provide two additional quantitative
measures of the fine-tuning in these same cases. We have also
included entry No.~10 from Table~\ref{tbl:signals} for comparison,
with high scale input values $\tan\beta=25$, $m_0 = 500 \GeV$,
$m_{1/2} = 300 \GeV$, $A_0 = -750\GeV$ and positive $\mu$ term.
The numbers $\delta_Z$ and $\delta_A$ are the sensitivities of
$M_Z$ and $m_A$, respectively, to small changes in the values of
the independent high-scale values $a_i$; i.e. $\delta = \sqrt{\sum
(\delta_i)^2}$ where $\delta_i = |(a_i/m) \Delta m/\Delta a_i |
$~\cite{BaGi88}. In order to treat unified and non-unified models
equally we have used the average scalar mass squared, gaugino mass
and trilinear coupling as free variables in computing these
sensitivities.

%================ Fine tuning table ==============================
\begin{table}[tb]
\caption{\label{tbl:tuning} Measures of fine tuning with respect
to high scale parameters in selected entries from
Table~\ref{tbl:signals}. For example, the entries for model~1
imply that a 1\% shift in high scale parameters leads to a 953\%
shift in the value of $m_A^2$.}
\begin{ruledtabular}
\begin{tabular}{|c|c|c|c|c|c|c|c|} \cline{1-8}
Entry & 1 & 3 & 4 \& 9 & 8 & 10 & 15 & 16 \\
\hline \hline
Parameters & 10 & 10 & 9 & $3^{a}$ & $3^{a}$ & 11 & $3^{a}$ \\
\cline{1-8}
$\delta_Z$ & 1003 & 363 & 1250 & 89 & 83 & 28600 & 135 \\
\cline{1-8}
$\delta_A$ & 953 & 135 & 640 & 80 & 1.4 & 275 & 93 \\
\cline{1-8}
\end{tabular}
\end{ruledtabular}
\begin{flushleft} $^{a}$ Does not include the specification of $\tan\beta$ and
the sign of the $\mu$ parameter.
\end{flushleft}
\end{table}
%===========================================================================

As far as we can see, all models with $m_A \sim m_h$, or
equivalently $C_H \sim 1$ are significantly fine-tuned. This is
not clear from the low-scale parameters, but seems to emerge when
one examines the high-scale models that give rise to small $m_A$.
Models which require specifying multiple soft parameters quite
precisely also imply additional tuning costs relative to the
mSUGRA models. This should be seen as evidence of the difficulty
in finding areas of the low-energy parameter space capable of
producing many of the entries in Table~\ref{tbl:signals}. While
the fine-tuning ``price'' of the LEP results for the MSSM has been
often discussed~\cite{ChElOlPo99}, it is apparent from
Table~\ref{tbl:tuning} that the least fine-tuned result continues
to be the case with $m_h \simeq 115\GeV$.

Yet even this outcome tends to require some superpartner masses
heavier that one might naively expect, in order to obtain the $(75
\GeV)^2$ radiative correction. In the standard mSUGRA-based
studies~\cite{ElHeOlWe03,ElNaOl02} one typically needs here either
squarks or gluinos in excess of $1 \TeV$ in mass at the low energy
scale, with the latter being a much more serious problem for
fine-tuning than the former~\cite{KaKi99,KaLyNeWa02}. Most of
these studies assume vanishing trilinear A-terms, however. The
degree of tuning can be reduced substantially if the so-called
``maximal mixing'' scenario can work~\cite{CaHeWaWe99}. In this
case, the need for large superpartner masses is mitigated by
maximizing the loop correction to the lightest Higgs boson mass
from the $m^{2}_{\rm LR}$ entry of the stop mass matrix. In models
whose scalar sector is well approximated by an overall universal
scalar mass $m_0$, this tends to occur when $A_t \simeq -2 m_0$ at
the GUT scale~\cite{KaKiWa01}. In models with small departures
from universality this relation remains approximately correct.

This relation is therefore an alluring goal for high-energy
models, though few well-motivated models seem to naturally predict
this relation without simultaneously re-introducing some heavy
superpartners, and often the hierarchies or negative mass-squareds
mentioned above still occur here. Minimal supergravity and minimal
anomaly mediation effectively treat both trilinears and scalar
masses as independent variables so no such relation is predicted.
Minimal gauge mediation predicts trilinears which are small
relative to scalar masses. In string-based examples some similar
relations are predicted. For example, the dilaton domination model
with tree-level K\"ahler potential~\cite{BrIbMu94} for the dilaton
provides a relation $A_0 = - \sqrt{3} m_0 \simeq -1.7 m_0$ at the
SUSY breaking scale. On the other hand, since $M_{1/2} = -A_0$ in
this model a sufficiently heavy Higgs state will require a gluino
mass that is also rather large. In the generalized dilaton
domination scenario~\cite{Ca96,BiGaWu97a,bench} gauginos masses
are parametrically small relative to scalars, but so too are the
trilinears. Allowing some coupling of the Higgs fields to the
Green-Schwarz counterterm in such theories can help generate
larger A-terms, but generally suffer from light moduli
problems~\cite{GaNe00a,GaGi00}.

In heterotic models where moduli fields participate in
supersymmetry breaking the tree-level trilinear A-terms vanish in
the limit where the Yukawa couplings are independent of the moduli
(as in the case with universal modular weights
$n=-1$)~\cite{BrIbMu94}. The relation $A \sim - m_0$ can be
obtained by taking larger (negative) modular weights, but this
introduces tachyonic scalar masses for at least some of the
fields. While non-negative modular weights are possible in
string-derived heterotic orbifold models~\cite{IbLu92}, which
could generate $A \sim - m_0$ with positive scalar mass-squareds,
such outcomes are hardly generic and require very special
conditions on the oscillator numbers of the fields involved.

Models based on intersecting D-brane
configurations~\cite{IbMuRi99} relate compactification moduli to
the dilaton through various duality symmetries. Thus a ``weight''
can be assigned to all moduli fields, including the dilaton. While
the soft trilinear A-terms continue to be given by the same
supergravity-derived expression, the effective weights of the MSSM
fields need no longer be integers. Depending on the brane
assignments of the Standard Model fields, relations such as $A_t =
- m_{\tilde{q}}$, -$\sqrt{2} m_{\tilde{q}}$ and $A_t = -\sqrt{3}
m_{\tilde{q}}$ can be obtained, with $m_{\tilde{q}}$ being the
typical squark mass at the high scale. Non-integer and
non-negative effective modular weights have recently been shown to
be possible in weakly-coupled heterotic constructions as well,
whenever MSSM fields are charged under an anomalous $U(1)$ factor
and fields which get large vevs to cancel the Fayet-Iliopolous
term are integrated out in a modular invariant
manner~\cite{GaGi02b}. Thus we see that the ability to imply a
result like $A_0 \simeq -2 m_0$ may be a significant discriminant
among string-based models and mechanisms of transmitting
supersymmetry breaking. A further analysis of many of these
string-based approached to achieving a large Higgs mass without
fine-tuning will be presented elsewhere~\cite{future}.

Given the above, the ``best'' answer for the MSSM remains that LEP
did produce 115 GeV Higgs states, and that this mass was the
result of some realization of what we will call the ``constrained
maximal mixing'' scenario, with $A_t \simeq -2 m_0$ and $M_3$,
$\mu$ small enough so that radiative electroweak symmetry breaking
occurs without large cancellations~\cite{KaLyNeWa02}. How to
generate this constrained maximal mixing scenario in realistic
high-scale theories is not clear, though some approaches may be
good starting points. For instance, we find it interesting that
the minimal SO(10) model of Raby et al. arrive at the same
relationship between the A-terms and squark masses as a necessary
condition for natural third-generation Yukawa
unification~\cite{Ra03,BlDeRa02}. Alternatively, perhaps the LEP
results suggest that what is needed is an extension to the usual
MSSM electroweak sector, such as an additional $U(1)$ sector whose
associated scalars mix strongly with the MSSM Higgs
sector~\cite{CvLa96,LaWa98,BaHuKiRoVe00}. Such an extension must
therefore arise at or near the weak scale. Perhaps the data is
pointing us towards a still (relatively) unexplored, but crucial,
region of theoretical parameter space whose theoretical simplicity
we cannot yet see.

\begin{acknowledgments}
The authors would like to thank S.~King and J.~D.~Wells for
helpful conversations and suggestions.
\end{acknowledgments}

%\bibliography{omega}

\end{document}